# Environmental Doping-Induced Degradation of the Quantum Anomalous Hall Insulators

Han Tay [1,3], Yi-Fan Zhao[1,3], Ling-Jie Zhou[1], Ruoxi Zhang[1], Zi-Jie Yan[1], Deyi Zhuo[1], Moses H. W. Chan[1], and Cui-Zu Chang[1,2]

[1] Department of Physics, The Pennsylvania State University, University Park, PA 16802, USA

[2] Materials Research Institute, The Pennsylvania State University, University Park, PA 16802, USA

[3] These authors contributed equally: Han Tay, Yi-Fan Zhao

Corresponding authors: cxc955@psu.edu (C.-Z. C.).

**Abstract:** **The quantum anomalous Hall (QAH) insulator is a topological quantum state with quantized Hall resistance and zero longitudinal resistance in the absence of an external magnetic field. The QAH insulator carries spin-polarized dissipation-free chiral edge current and thus provides a unique opportunity to develop energy-efficient transformative information technology. Despite promising advances on the QAH effect over the past decade, the QAH insulator has thus far eluded any practical applications. In addition to its low working temperature, the QAH state in magnetically doped topological insulator (TI) films/heterostructures usually deteriorates with time in ambient conditions. In this work, we prepare three QAH devices with similar initial properties and store them in different environments to investigate the evolution of their transport properties. The QAH device without a protection layer in air show clear degradation and becomes hole-doped with the charge neutral point shifting significantly to positive gate voltages. The QAH** device kept in an argon glove box without a protection layer shows no measurable degradation after 560 hours and the device protected by a 3 nm $AlO_x$ protection layer in air shows minimal degradation with stable QAH properties. Our work shows a route to preserve the dissipation-free chiral edge state in QAH devices for potential applications in quantum information technology.

**Keywords:** Quantum anomalous Hall insulator, topological insulator, molecular beam epitaxy growth, environmental doping, sample degradation

**Main text:** Three-dimensional (3D) topological insulators (TIs) possess conducting helical Dirac states on their surfaces but are electrically insulating in the interior. The Dirac surface states of TI are protected by time-reversal symmetry (TRS) [1,2]. By introducing a ferromagnetic order to break TRS, a magnetic exchange gap is created at the Dirac point, leading to a quantized topological phenomenon known as the quantum anomalous Hall (QAH) effect [3-8]. In the past 15 years, hundreds of materials have been theoretically predicted and experimentally demonstrated as TIs (Refs. [9-11]). Among these TIs, the most appealing TI material is the $Bi_2Se_3$ family, which includes three stoichiometric members: $Bi_2Se_3$, $Bi_2Te_3$, and $Sb_2Te_3$ (Refs. [12-15]). These three TIs and their compounds have a large bulk gap of 260~300 meV and a single Dirac cone at the $\Gamma$ point of the two-dimensional (2D) surface Brillion zone. Since the vacancies (e.g., Se and Te) and anti-sites [e.g., Bi/(Te,Se) and Sb/Te] inevitably exist, the $Bi_2Se_3$ family TIs are usually electron- or hole-doped. In other words, it is challenging to single out the contribution of the helical surface states through electrical transport measurements. Moreover, the $Bi_2Se_3$ family TIs are exceptionally sensitive to ambient conditions [16-23]. Air consists of $N_2$ (~78%), $O_2$ (~21%), water vapor ($H_2O$, variable) and other trace gases. $N_2$ is chemically inert and thus unlikely to degrade TIs. Prior studies have shown that $O_2$ introduces hole-type carriers and shifts the chemical potential downward upon the occupation of Se/Te vacancies [16,18]. However, water vapor introduces electron-type carriers and shifts the chemical potential upward because of the formations of the positively charged Se/Te vacancies and bismuth hydroxide [17].

Through the diluted magnetic doping approach, the QAH effect has been realized in two magnetically doped TI systems, i.e., Cr- and/or V-doped (Bi,Sb)$_2$Te$_3$ (Refs. [3, 8, 24-42]). These two systems share a parent TI material, i.e., (Bi,Sb)$_2$Te$_3$, in which the Bi/Sb ratio tunes the chemical potential location [43, 44]. When the chemical potential is tuned into the magnetic exchange gap at the Dirac point, the magnetically doped TI films/heterostructures become a QAH insulator [3, 8, 24-42]. In real experiments, it is known that exposure to air shifts the chemical potential of the magnetically doped TI films/heterostructures. Since the size of the effective magnetic exchange gap is only a few meV (Refs. [3, 36, 45]), the property of the QAH state is sensitive to ambient conditions. To date, the QAH effect has been realized in magnetically doped TI thin films/heterostructures either without [8, 30-35] or with [24-26, 28, 34-42] protection layers. The protection layers include amorphous Te (Refs. [24, 35-37, 39]) and aluminum [26, 34, 38, 46] films grown by *in situ* molecular beam epitaxy (MBE) and an AlO$_x$ layer [25, 28, 40-42] grown by *ex situ* atomic layer deposition (ALD). For the *in situ* grown amorphous Te and aluminum films, both may turn out to be corresponding oxides after the QAH samples are taken out of the MBE chamber [24, 26, 34-39, 46]. So far, the evolution of the QAH transport properties under ambient environments has not been studied and the function of the protection layers is unclear.

In this work, based on our MBE-grown QAH sandwich heterostructures, specifically 3 quintuple layers (QL) (Bi, Sb)$_{1.78}$Cr$_{0.22}$Te$_3$/8 QL (Bi, Sb)$_2$Te$_3$/3 QL (Bi, Sb)$_{1.78}$Cr$_{0.22}$Te$_3$ (Supplementary Fig. 1)(Refs. [30-32]), we prepare three QAH devices with similar initial transport properties and keep them under different environments. The evolution of their electrical transport properties over 500 hours is then periodically probed. For the QAH device without a protection layer stored in air (Device A, Fig. 2a), we find that the ratio between its Hall resistance and longitudinal resistance at zero magnetic field first increases and then gradually decreases, and its charge neutral point shifts to larger positive gate voltages, indicating the introduction of hole-type carriers to the QAH device in air. For the QAH device with a 3 nm AlO$_x$ protection layer in air (Device B, Fig. 2b) and the QAH device without a protection layer in an argon glove box (Device C, Fig. 2c), the QAH state is slightly improved (Device B) or shows no noticeable change (Device C) over 500 hours. Therefore, adding an AlO$_x$ protection layer mimics the effect of storing the QAH device in an argon glove box. Both approaches can significantly reduce the environmental doping-induced hole-type carriers and thus mitigate the degradation of the QAH insulators.

The QAH sandwiches are grown on heat-treated ~0.5 mm thick SrTiO$_3$(111) substrates in a commercial MBE system (Omicron Lab 10) with a base vacuum better than $1\times10^{-10}$ mbar (Refs. [30-32]). Next, we scratch these QAH samples into a Hall bar geometry with an effective area of ~0.5 mm × 1.0 mm (Figs. 1a and 1b). Both the electrical contacts and the bottom gate electrode are made by pressing indium dots. The electrical transport measurements are conducted in a Physical Property Measurement System (PPMS, Quantum Design DynaCool, 1.7 K, 9 T) and a Leiden Cryogenics dilution refrigerator (10 mK, 9 T) with the magnetic field applied perpendicular to the film plane. More details about the MBE growth, device fabrications, sample characterizations, and electrical transport measurements can be found in Supporting Information.

Before we investigate the degradation of the QAH insulators under different ambient environments, we first perform dilution transport measurements on a typical magnetically doped TI sandwich without a protection layer to characterize the QAH effect. This sample is exposed to air for a total time of less than 2 hours before being loaded into the vacuum chamber of the dilution refrigerator and cooled down to 4 K (Figs. 1a and 1b). At $T = 25$ mK and the charge neutral point $V_g = V_g^0$, the sample shows the well-quantized QAH effect (Fig. 1c). The Hall resistance $\rho_{yx}$ at zero magnetic field [labeled as $\rho_{yx}(0)$] is found to be ~0.991 $h/e^2$, concomitant with the longitudinal resistance $\rho_{xx}$ at zero magnetic field [labeled as $\rho_{xx}(0)$] ~$0.0008h/e^2$ (~20 Ω) (Fig.1c). To further validate the appearance of the QAH effect, we measure magnetic field $\mu_0H$ dependence of $\rho_{yx}$ and $\rho_{xx}$ at different $V_g$s (Supplementary Fig. 2). Figure 1d shows the $V_g$ dependence of $\rho_{yx}(0)$ and $\rho_{xx}(0)$. We find that $\rho_{yx}(0)$ exhibits a quantized plateau of ~$h/e^2$ and the corresponding $\rho_{xx}(0)$ shows a zero-resistance plateau near $V_g = V_g^0$. This observation confirms the appearance of the well-quantized QAH effect in our magnetically doped TI sandwiches. For $V_g>V_g^0$, the chemical potential first crosses the helical surface states, which introduces few carriers, and thus the quantized $\rho_{yx}(0)$ plateau preserves and extends to large positive gate voltages. However, for $V_g<V_g^0$, the chemical potential first crosses the bulk valence bands, and thus plenty of carriers are introduced, which leads to a large deviation from the QAH state [3, 36, 40, 45, 47].

Based on our MBE-grown QAH sandwich heterostructures, which usually show very similar QAH behavior, specifically similar $\rho_{yx}$ and $\rho_{xx}$ at $V_g = V_g^0$, we fabricate three Hall bar devices and store them under different ambient environments. Device A without a protection layer is placed in air, which is exposed to $N_2$, $O_2$, and water vapor directly (Fig. 2a). Device B with a 3 nm $AlO_x$ protection layer grown by *ex situ* ALD is also placed in air (Fig. 2b). Device C without a protection layer is placed in an argon glove box of less than 0.1 parts per million (ppm) of both $O_2$ and $H_2O$ (Fig. 2c). For each device, the initial sample preparation (e.g., pressing electrical contacts and mounting the device onto the sample puck) takes less than 30 minutes after its removal from the MBE chamber and each measurement under rough vacuum at different storage time $t$ takes less than 40 minutes. Device C is exposed to air for less than 5 minutes for each subsequent measurement at different storage time $t$. Next, we carry out transport measurements at $T$~1.7 K in a PPMS cryostat with a short turnaround time to track the evolution of the QAH state every 6~8 hours for a total storage time $t$ of 500~560 hours (Supplementary Figs. 3 and 4).

Figures 2d and 2g show the magnetic field $\mu_0 H$-dependent $\rho_{yx}$ and $\rho_{xx}$ of Device A at different storage time $t$. At $t = 0$ h, $\rho_{yx}(0) \sim 0.880\ h/e^2$ and $\rho_{xx}(0) \sim 0.268\ h/e^2$, which corresponds to a Hall angle $\alpha \sim 73.1°$. With increasing storage time $t$, $\rho_{yx}(0)$ first increases and then gradually decreases (Fig. 2g). The values of $\rho_{yx}(0)$ are $\sim 0.918\ h/e^2$, $\sim 0.839\ h/e^2$, and $\sim 0.830\ h/e^2$ for $t = 20$ h, $t = 210$ h, and $t = 500$ h, respectively (Fig. 2g). The corresponding $\rho_{xx}(0)$ are $\sim 0.207\ h/e^2$, $\sim 0.347\ h/e^2$, and $\sim 0.365\ h/e^2$ (Fig. 2d). The Hall angle $\alpha \sim 77.3°$, $\sim 67.5°$, and $66.3°$ for $t = 20$ h, $t = 210$ h, and $t = 500$ h, respectively. For Device B, with the increase of $t$, $\rho_{yx}(0)$ increases gently (Fig. 2h). The values of $\rho_{yx}(0)$ are $\sim 0.876\ h/e^2$, $\sim 0.900\ h/e^2$, $\sim 0.927\ h/e^2$, and $\sim 0.933\ h/e^2$ for $t = 0$ h, $t = 25$ h, $t = 250$ h, and $t = 560$ h, respectively (Fig. 2h). The corresponding $\rho_{xx}(0)$ are $\sim 0.319\ h/e^2$, $\sim 0.224\ h/e^2$, $\sim 0.151\ h/e^2$, and $\sim 0.136\ h/e^2$ (Fig. 2e). The Hall angle $\alpha \sim 70.0°$, $\sim 76.0°$, $\sim 80.7°$, and $\sim 81.7°$ for $t = 0$ h, $t = 25$ h, $t = 250$ h, and $t = 560$ h, respectively. The QAH sample slightly degrades during the deposition of the 3 nm $AlO_x$ protection layer by *ex situ* ALD, but the sample returns to a better QAH state in air for a few days. For Device C, with the increase of $t$, $\rho_{yx}(0)$ does not show pronounced change (Fig. 2i). The values of $\rho_{yx}(0)$ are $\sim 0.871\ h/e^2$, $\sim 0.865\ h/e^2$, $\sim 0.873\ h/e^2$, and $\sim 0.879\ h/e^2$ for $t = 0$ h, $t = 70$ h, $t = 240$ h, and $t = 560$ h, respectively (Fig. 2i). The corresponding $\rho_{xx}(0)$ are $\sim 0.220\ h/e^2$, $\sim 0.238\ h/e^2$, $\sim 0.224\ h/e^2$, and $\sim 0.215\ h/e^2$ (Fig. 2f). The Hall angle $\alpha \sim 75.8°$, $\sim 74.6°$, $\sim 75.6°$, and $\sim 76.3°$ for $t = 0$ h, $t = 70$ h, $t = 240$ h, and $t = 560$ h, respectively. We note that the different values of the $\rho_{xx}$ near the coercive fields $\mu_0 H_c$ might signify the different QAH sample degradations under different ambient environments.

To reveal more clearly the evolution of the environmental doping-induced degradation of the QAH insulators, we plot the $\rho_{yx}(0)$ and $\rho_{xx}(0)$ values of three QAH devices as a function of the storage time $t$ (Figs. 3a to 3c). For Device A, the QAH state first slightly improves and then gradually degrades. The maximum of $\rho_{yx}(0)$ ~0.918 $h/e^2$ appears at $t$ ~20h, concomitant with the minimum of $\rho_{xx}(0)$ ~0.207 $h/e^2$ (Fig. 3a), showing the largest Hall angle $\alpha$~77.3° (Fig. 3d). For Device B, the value of $\rho_{yx}(0)$ [$\rho_{xx}(0)$] monotonically increases (decreases). As noted above, $\rho_{yx}(0)$ ~0.876 $h/e^2$ and $\rho_{xx}(0)$ ~0.319 $h/e^2$ at $t$ ~0 h changes to $\rho_{yx}(0)$ ~0.933 $h/e^2$ and $\rho_{xx}(0)$ ~0.136 $h/e^2$ at $t$ ~560 h (Fig. 3b). The value of Hall angle $\alpha$ increases from ~70.0° at $t$ =0 h to ~81.7° at $t$ =560 h, respectively (Fig. 3d). For Device C, the values of $\rho_{yx}(0)$ and $\rho_{yx}(0)$ show a weak fluctuation through the duration of our experiment. The largest $\rho_{yx}(0)$ ~ 0.882 $h/e^2$ and smallest $\rho_{xx}(0)$ ~0.210 $h/e^2$ are found at t~ 480 h, while the smallest $\rho_{yx}(0)$ ~0.863 $h/e^2$ and largest $\rho_{xx}(0)$ ~0.241 $h/e^2$ are found at $t$ ~50h(Fig. 3c). The Hall angle $\alpha$ fluctuates within a small range from ~75.8° at $t$=0 h to ~74.4° at $t$=50 h to ~76.3° at $t$=560 h. (Fig. 3d).

We note that the degradation of Devices A and B occurs in the first 200 to 250 hours. For the QAH device without a protection layer in air (Device A), its quality reaches its peak within a short time of ~20h but then gradually degrades with $t$. Since the $\rho_{yx}(0)/\rho_{xx}(0)$ of Device A is still ~2.274 at $t$~500 h, much greater than 1, suggesting that the QAH properties persist after ~20 days in air without a protection layer (Fig. 3a)[3, 31]. The 3 nm AlO$_x$ protection layer (Device B) appears to be nearly effective in preserving the QAH state as storing a device in an argon glove box (Device C) (Figs. 3b to 3d).

Next, we investigate the shift of the charge neutral point $V_g^0$ over the storage time $t$ for each of the three devices. To achieve the value of $V_g^0$ at different storage time $t$, we measure the $V_g$ dependence of $\rho_{yx}$ and $\rho_{xx}$ at $\mu_0 H$ =1 T and $T$ =1.7 K (Figs. 4a to 4f). We observe clear ambipolar features in $V_g$-dependent $\rho_{yx}$ and $\rho_{xx}$ curves [44, 48]. The value of $V_g^0$ is approximately determined as the gate voltage $V_g$ where $\rho_{yx}$ is maximized [3]. For Device A, the $V_g^0$ value monotonically increases with increasing the storage time $t$. $V_g^0 \sim 0$ V at $t=0$h shifts to $V_g^0 \sim +141$ V at $t=500$ h (Figs. 4a, 4b, and 4g). Therefore, the hole-type carriers are introduced to the QAH sample, which is probably a result of the occupation of Te in the QAH sample by the $O_2$ in air [16, 18, 20]. It appears that the water vapor in air may play a negligible role in the degradation of the QAH insulators. For Device B, a similar but much smaller $V_g^0$ shift is observed. $V_g^0 \sim -16$V at $t=0$ h shifts to $V_g^0 \sim +11$ V at $t=560$ h (Figs. 4c, 4d, and 4h). The reduced $V_g^0$ shift indicates the 3 nm $AlO_x$ protection layer effectively protects the QAH sample against environmental-doping induced-degradation. As noted above, the QAH quantization becomes slightly better after being exposed to air for tens of days. For Device C, the $V_g^0$ value is nearly pinned. $V_g^0$ is found $\sim -12$ V at $t = 0$ h and $\sim -14$ V at $t =560$ h (Figs. 4e, 4f, and 4i). This confirms the practice that the QAH device without a protection layer can be safely protected in an argon glove box.

To summarize, we investigate the environmental doping-induced degradation of the QAH insulators. We find that exposure to air indeed degrades the sample quality of the QAH samples. Overall, air introduces hole-type carriers to the QAH insulators, which drive the magnetically doped TI films/heterostructures away from the QAH state. By storing the QAH device in an argon glove box, no visible hole-type carriers are introduced and thus the properties of the QAH insulators are effectively protected. By adding a 3 nm $AlO_x$ protection layer, the air-induced hole-type doping is greatly reduced and the properties of the QAH insulators are also effectively protected. Since most of the electrical transport measurements and device fabrication processes are conducted in ambient environments, a comprehensive understanding of the air-induced degradation of the QAH insulators and the establishment of stable methods for the long-time storage of QAH insulators are crucial for the future development of the energy-efficient QAH-based quantum information technology [3].

**Supporting Information.** The Supporting Information is available free of charge on the ACS Publications website.

MBE growth of the QAH sandwich heterostructures, QAH Hall bar device fabrication, electrical transport measurements, reflection high-energy electron diffraction (RHEED) patterns of the QAH sample, and more transport results of the QAH devices.

**Author contributions:** C.-Z. C. conceived and supervised the experiment. Y.-F. Z., Z.-J. Y., and D. Z. grew the magnetically doped TI sandwiches with the QAH state. L.-J. Z deposited the $AlO_x$ capping layer on the QAH devices. H. T. and Y.-F. Z. carried out the PPMS transport measurements. R. Z. performed the dilution transport measurements. H. T., Y.-F. Z., and C.-Z. C. analyzed the data and wrote the manuscript with input from all authors.

**Notes:** The authors declare no competing financial interest.

**Acknowledgments:** We thank C.-X. Liu and N. Samarth for helpful discussions. This work is primarily supported by the ARO Award (W911NF2210159), including MBE growth, device fabrication, and dilution transport measurements. The PPMS measurements are partially supported by the AFOSR grant (FA9550-21-1-0177). The deposition of the $AlO_x$ protection layer and the data analysis are partially supported by the DOE grant (DE-SC0023113). C. Z. C. also acknowledges the support from the Gordon and Betty Moore Foundation's EPiQS Initiative (Grant GBMF9063 to C.-Z. C.).

**Figures and figure captions:**

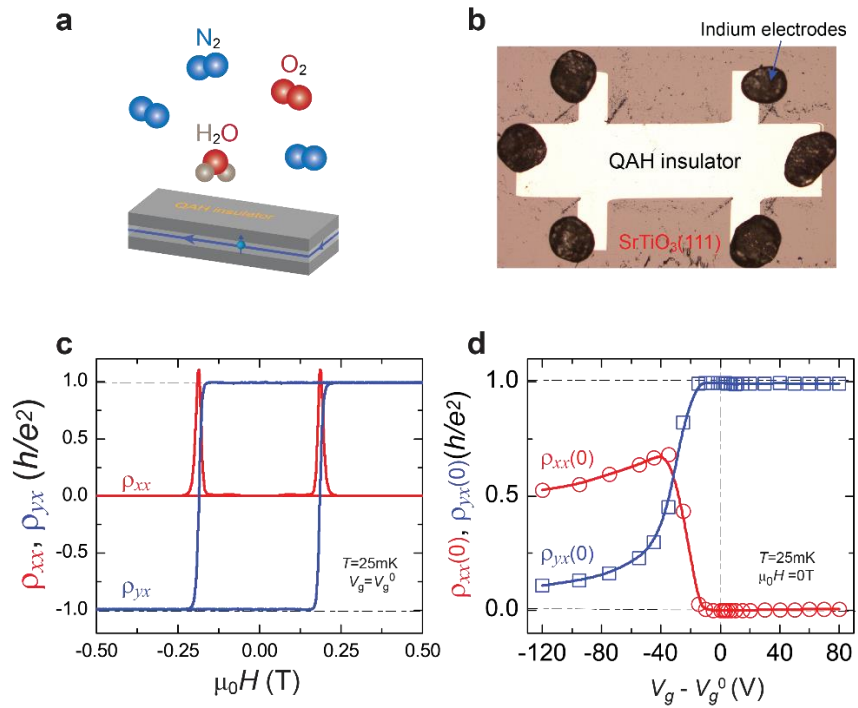

**Fig. 1| The QAH effect observed in a magnetically doped TI sandwich without a protection layer. a,** Schematic of the QAH device, specifically 3 QL Cr-doped $(Bi,Sb)_2Te_3$/8 QL $(Bi,Sb)_2Te_3$/3 QL Cr-doped $(Bi,Sb)_2Te_3$, under ambient conditions with $N_2$, $O_2$, and water vapor ($H_2O$). **b,** The optical photograph of the QAH Hall bar device used in transport measurements. The effective area of the Hall bar is ~1.0mm × 0.5mm. **c,** $\mu_0H$ dependence of $\rho_{xx}$ (red) and $\rho_{yx}$ (blue) measured at $V_g = V_g^0$ and $T = 25$ mK. **d,** Gate ($V_g - V_g^0$) dependence of $\rho_{xx}(0)$ (red circles) and $\rho_{yx}(0)$ (blue squares) measured at $T = 25$ mK and $\mu_0H = 0$ T. The QAH Hall bar device used in (c) and (d) is exposed to air for less than 2 hours.

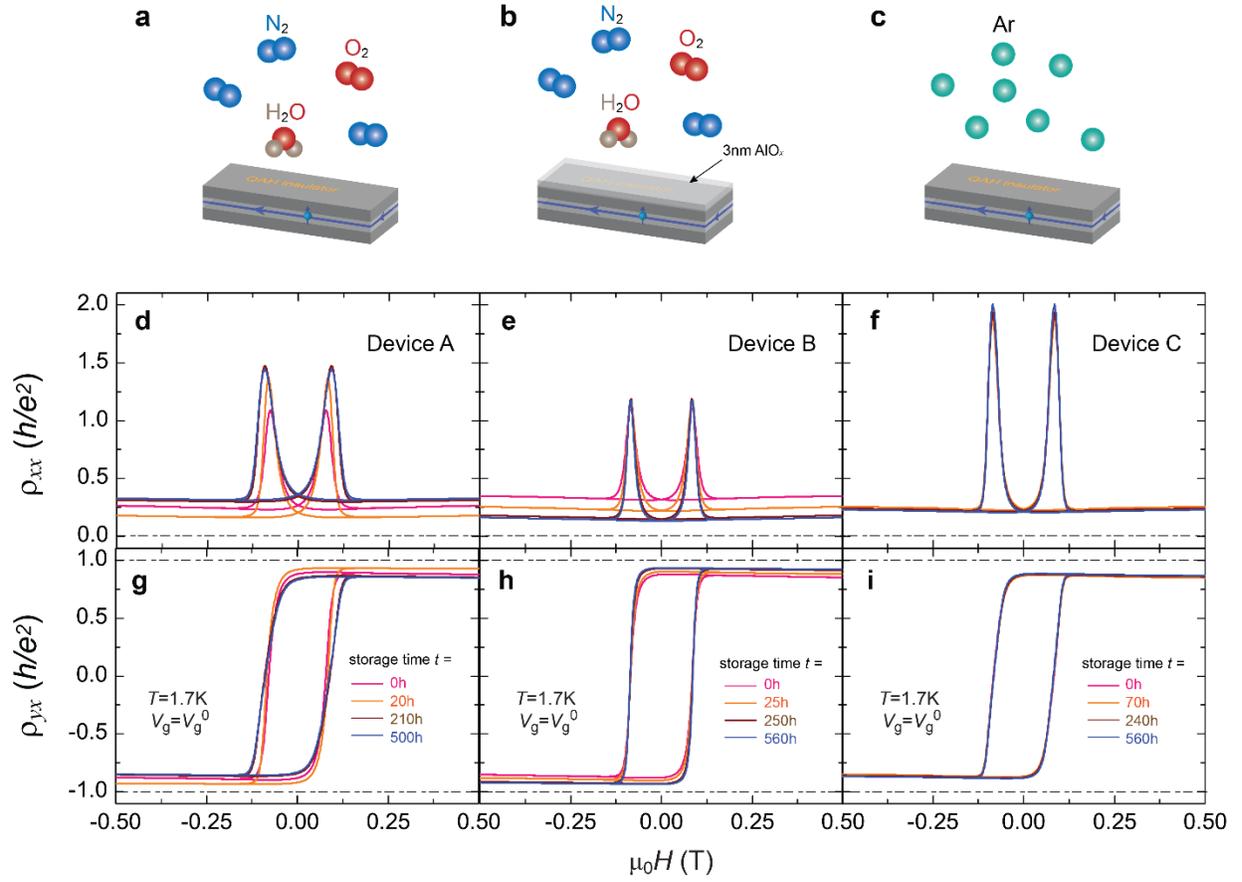

**Fig. 2| Degradation of the QAH insulators under different ambient environments. a-c,** Schematics of the QAH insulators under different circumstances. (a) Device A without a protection layer in air; (b) Device B with a 3 nm AlO$_x$ protection layer in air; (c) Device C without a protection layer in an argon glove box. **d-f**, $\mu_0 H$ dependent $\rho_{xx}$ of the three QAH devices measured at different storage times. **g-i**, Same as (d-f), but for $\rho_{yx}$. All measurements are performed at $V_g = V_g^0$ and $T$ =1.7 K.

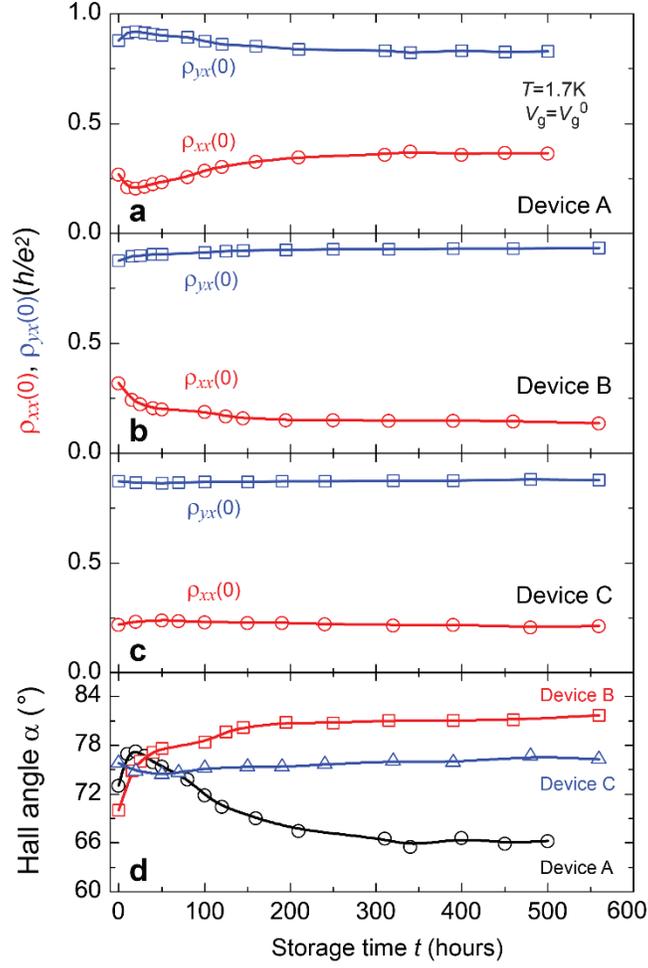

**Fig. 3| Evolution of the environmental doping-induced QAH degradation. a-c**, Time-dependent $\rho_{xx}(0)$ (red circles) and $\rho_{yx}(0)$ (blue squares) of Device A (a), Device B (b), and Device C (c). **d**, Time-dependent Hall angle $\alpha$ of Device A (black circles), Device B (red squares), and Device C (blue triangles).

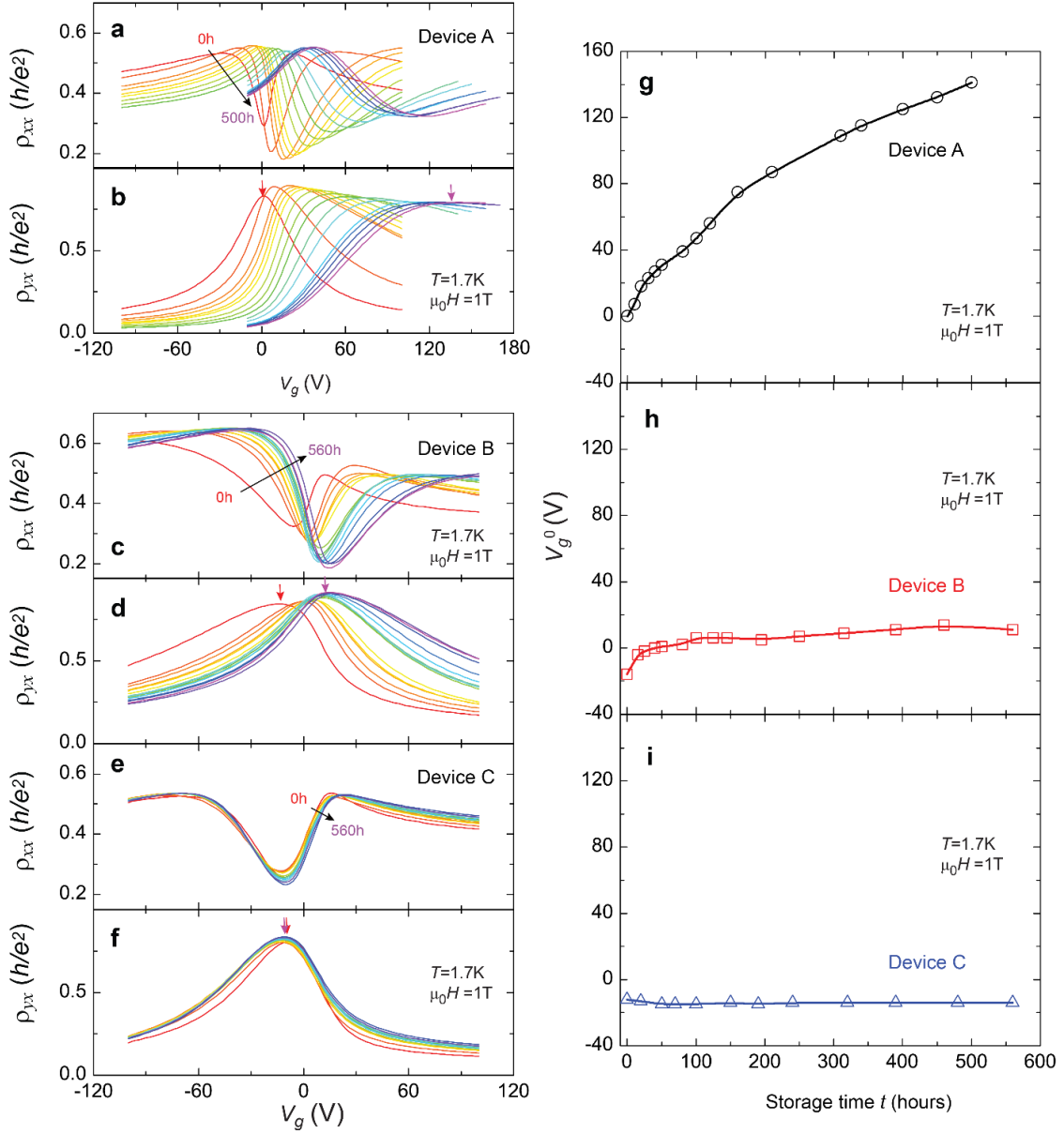

**Fig. 4| Environmental doping-induced charge neutral point shift in QAH insulators. a-f**, $V_g$ dependent $\rho_{xx}$ and $\rho_{yx}$ of Device A (a,b), Device B (c,d), and Device C (e,f) measured at different times. The red and purple arrows mark the values of the charge neutral point $V_g^0$ at 0h and 500h for Device A, 0h and 560h for Devices B and C. **g-i**, Storage time $t$-dependent $V_g^0$ values of Device A (black circles, g), Device B (red squares, h), and Device C (blue triangles, i). All measurements are performed at $T = 1.7$ K and $\mu_0 H = 1$ T.

**References:**

1. Hasan, M. Z.; Kane, C. L. Colloquium: Topological Insulators. *Rev. Mod. Phys.* **2010,** 82, 3045-3067.
2. Qi, X. L.; Zhang, S. C. Topological Insulators and Superconductors. *Rev. Mod. Phys.* **2011,** 83, 1057-1110.
3. Chang, C.-Z.; Liu, C.-X.; MacDonald, A. H. Colloquium: Quantum anomalous Hall effect. *arXiv:2202.13902* **2022**.
4. Haldane, F. D. M. Model for a Quantum Hall-Effect without Landau Levels: Condensed-Matter Realization of the "Parity Anomaly". *Phys. Rev. Lett.* **1988,** 61, 2015-2018.
5. Liu, C. X.; Qi, X. L.; Dai, X.; Fang, Z.; Zhang, S. C. Quantum anomalous Hall effect in $Hg_{1-y}Mn_yTe$ quantum wells. *Phys. Rev. Lett.* **2008,** 101, 146802.
6. Qi, X. L.; Hughes, T. L.; Zhang, S. C. Topological Field Theory of Time-Reversal Invariant Insulators. *Phys. Rev. B* **2008,** 78, 195424.
7. Yu, R.; Zhang, W.; Zhang, H. J.; Zhang, S. C.; Dai, X.; Fang, Z. Quantized Anomalous Hall Effect in Magnetic Topological Insulators. *Science* **2010,** 329, 61-64.
8. Chang, C. Z.; Zhang, J. S.; Feng, X.; Shen, J.; Zhang, Z. C.; Guo, M. H.; Li, K.; Ou, Y. B.; Wei, P.; Wang, L. L.; Ji, Z. Q.; Feng, Y.; Ji, S. H.; Chen, X.; Jia, J. F.; Dai, X.; Fang, Z.; Zhang, S. C.; He, K.; Wang, Y. Y.; Lu, L.; Ma, X. C.; Xue, Q. K. Experimental Observation of the Quantum Anomalous Hall Effect in a Magnetic Topological Insulator. *Science* **2013,** 340, 167-170.
9. Zhang, T. T.; Jiang, Y.; Song, Z. D.; Huang, H.; He, Y. Q.; Fang, Z.; Weng, H. M.; Fang, C. Catalogue of topological electronic materials. *Nature* **2019,** 566, 475-479.
10. Vergniory, M. G.; Elcoro, L.; Felser, C.; Regnault, N.; Bernevig, B. A.; Wang, Z. J. A complete catalogue of high-quality topological materials. *Nature* **2019,** 566, 480-485.
11. Tang, F.; Po, H. C.; Vishwanath, A.; Wan, X. G. Comprehensive search for topological materials using symmetry indicators. *Nature* **2019,** 566, 486-489.
12. Zhang, H. J.; Liu, C. X.; Qi, X. L.; Dai, X.; Fang, Z.; Zhang, S. C. Topological Insulators in $Bi_2Se_3$, $Bi_2Te_3$ and $Sb_2Te_3$ with a Single Dirac Cone on the Surface. *Nat. Phys.* **2009,** 5, 438-442.
13. Xia, Y.; Qian, D.; Hsieh, D.; Wray, L.; Pal, A.; Lin, H.; Bansil, A.; Grauer, D.; Hor, Y. S.; Cava, R. J.; Hasan, M. Z. Observation of a Large-Gap Topological-Insulator Class with a Single Dirac Cone on the Surface. *Nat. Phys.* **2009,** 5, 398-402.

14. Chen, Y. L.; Analytis, J. G.; Chu, J. H.; Liu, Z. K.; Mo, S. K.; Qi, X. L.; Zhang, H. J.; Lu, D. H.; Dai, X.; Fang, Z.; Zhang, S. C.; Fisher, I. R.; Hussain, Z.; Shen, Z. X. Experimental Realization of a Three-Dimensional Topological Insulator, $Bi_2Te_3$. *Science* **2009,** 325, 178-181.

15. Hsieh, D.; Xia, Y.; Qian, D.; Wray, L.; Meier, F.; Dil, J. H.; Osterwalder, J.; Patthey, L.; Fedorov, A. V.; Lin, H.; Bansil, A.; Grauer, D.; Hor, Y. S.; Cava, R. J.; Hasan, M. Z. Observation of Time-Reversal-Protected Single-Dirac-Cone Topological-Insulator States in $Bi_2Te_3$ and $Sb_2Te_3$. *Phys. Rev. Lett.* **2009,** 103, 146401.

16. Chen, Y. L.; Chu, J. H.; Analytis, J. G.; Liu, Z. K.; Igarashi, K.; Kuo, H. H.; Qi, X. L.; Mo, S. K.; Moore, R. G.; Lu, D. H.; Hashimoto, M.; Sasagawa, T.; Zhang, S. C.; Fisher, I. R.; Hussain, Z.; Shen, Z. X. Massive Dirac Fermion on the Surface of a Magnetically Doped Topological Insulator. *Science* **2010,** 329, 659-662.

17. Benia, H. M.; Lin, C. T.; Kern, K.; Ast, C. R. Reactive Chemical Doping of the $Bi_2Se_3$ Topological Insulator. *Phys. Rev. Lett.* **2011,** 107, 177602.

18. Koleini, M.; Frauenheim, T.; Yan, B. H. Gas Doping on the Topological Insulator $Bi_2Se_3$ Surface. *Phys. Rev. Lett.* **2013,** 110, 016403.

19. Hoefer, K.; Becker, C.; Rata, D.; Swanson, J.; Thalmeier, P.; Tjeng, L. H. Intrinsic conduction through topological surface states of insulating Bi2Te3 epitaxial thin films. *Proc. Natl. Acad. Sci.* **2014,** 111, 14979-14984.

20. Ngabonziza, P.; Heimbuch, R.; de Jong, N.; Klaassen, R. A.; Stehno, M. P.; Snelder, M.; Solmaz, A.; Ramankutty, S. V.; Frantzeskakis, E.; van Heumen, E.; Koster, G.; Golden, M. S.; Zandvliet, H. J. W.; Brinkman, A. In situ spectroscopy of intrinsic $Bi_2Te_3$ topological insulator thin films and impact of extrinsic defects. *Phys. Rev. B* **2015,** 92, 035405.

21. Brahlek, M.; Kim, Y. S.; Bansal, N.; Edrey, E.; Oh, S. Surface versus bulk state in topological insulator $Bi_2Se_3$ under environmental disorder. *Appl. Phys. Lett.* **2011,** 99, 012109.

22. Aguilar, R. V.; Wu, L.; Stier, A. V.; Bilbro, L. S.; Brahlek, M.; Bansal, N.; Oh, S.; Armitage, N. P. Aging and reduced bulk conductance in thin films of the topological insulator $Bi_2Se_3$. *J. Appl. Phys.* **2013,** 113, 153702.

23. Kong, D. S.; Cha, J. J.; Lai, K. J.; Peng, H. L.; Analytis, J. G.; Meister, S.; Chen, Y. L.; Zhang, H. J.; Fisher, I. R.; Shen, Z. X.; Cui, Y. Rapid Surface Oxidation as a Source of Surface

Degradation Factor for Bi$_2$Se$_3$. *ACS Nano* **2011,** 5, 4698-4703.

24. Chang, C. Z.; Zhao, W. W.; Kim, D. Y.; Zhang, H. J.; Assaf, B. A.; Heiman, D.; Zhang, S. C.; Liu, C. X.; Chan, M. H. W.; Moodera, J. S. High-Precision Realization of Robust Quantum Anomalous Hall State in a Hard Ferromagnetic Topological Insulator. *Nat. Mater.* **2015,** 14, 473-477.

25. Checkelsky, J. G.; Yoshimi, R.; Tsukazaki, A.; Takahashi, K. S.; Kozuka, Y.; Falson, J.; Kawasaki, M.; Tokura, Y. Trajectory of the Anomalous Hall Effect towards the Quantized State in a Ferromagnetic Topological Insulator. *Nat. Phys.* **2014,** 10, 731-736.

26. Kou, X. F.; Guo, S. T.; Fan, Y. B.; Pan, L.; Lang, M. R.; Jiang, Y.; Shao, Q. M.; Nie, T. X.; Murata, K.; Tang, J. S.; Wang, Y.; He, L.; Lee, T. K.; Lee, W. L.; Wang, K. L. Scale-Invariant Quantum Anomalous Hall Effect in Magnetic Topological Insulators beyond the Two-Dimensional Limit. *Phys. Rev. Lett.* **2014,** 113, 137201.

27. Ou, Y.; Liu, C.; Jiang, G.; Feng, Y.; Zhao, D.; Wu, W.; Wang, X. X.; Li, W.; Song, C.; Wang, L. L.; Wang, W.; Wu, W.; Wang, Y.; He, K.; Ma, X. C.; Xue, Q. K. Enhancing the Quantum Anomalous Hall Effect by Magnetic Codoping in a Topological Insulator. *Adv. Mater.* **2017,** 30, 1703062.

28. Mogi, M.; Yoshimi, R.; Tsukazaki, A.; Yasuda, K.; Kozuka, Y.; Takahashi, K. S.; Kawasaki, M.; Tokura, Y. Magnetic Modulation Doping in Topological Insulators toward Higher-Temperature Quantum Anomalous Hall Effect. *Appl. Phys. Lett.* **2015,** 107, 182401.

29. Chang, C. Z.; Zhang, J. S.; Liu, M. H.; Zhang, Z. C.; Feng, X.; Li, K.; Wang, L. L.; Chen, X.; Dai, X.; Fang, Z.; Qi, X. L.; Zhang, S. C.; Wang, Y. Y.; He, K.; Ma, X. C.; Xue, Q. K. Thin Films of Magnetically Doped Topological Insulator with Carrier-Independent Long-Range Ferromagnetic Order. *Adv. Mater.* **2013,** 25, 1065-1070.

30. Jiang, J.; Xiao, D.; Wang, F.; Shin, J. H.; Andreoli, D.; Zhang, J. X.; Xiao, R.; Zhao, Y. F.; Kayyalha, M.; Zhang, L.; Wang, K.; Zang, J. D.; Liu, C. X.; Samarth, N.; Chan, M. H. W.; Chang, C. Z. Concurrence of quantum anomalous Hall and topological Hall effects in magnetic topological insulator sandwich heterostructures. *Nat. Mater.* **2020,** 19, 732-737.

31. Zhao, Y. F.; Zhang, R.; Mei, R.; Zhou, L. J.; Yi, H.; Zhang, Y. Q.; Yu, J.; Xiao, R.; Wang, K.; Samarth, N.; Chan, M. H. W.; Liu, C. X.; Chang, C. Z. Tuning the Chern number in quantum anomalous Hall insulators. *Nature* **2020,** 588, 419-423.

32. Zhao, Y.-F.; Zhang, R.; Zhou, L.-J.; Mei, R.; Yan, Z.-J.; Chan, M. H. W.; Liu, C.-X.; Chang,

C.-Z. Zero Magnetic Field Plateau Phase Transition in Higher Chern Number Quantum Anomalous Hall Insulators. *Phys. Rev. Lett.* **2022,** 128, 216801.

33. Feng, Y.; Feng, X.; Ou, Y. B.; Wang, J.; Liu, C.; Zhang, L. G.; Zhao, D. Y.; Jiang, G. Y.; Zhang, S. C.; He, K.; Ma, X. C.; Xue, Q. K.; Wang, Y. Y. Observation of the Zero Hall Plateau in a Quantum Anomalous Hall Insulator. *Phys. Rev. Lett.* **2015,** 115, 126801.

34. Kandala, A.; Richardella, A.; Kempinger, S.; Liu, C. X.; Samarth, N. Giant anisotropic magnetoresistance in a quantum anomalous Hall insulator. *Nat. Commun.* **2015,** 6, 7434.

35. Winnerlein, M.; Schreyeck, S.; Grauer, S.; Rosenberger, S.; Fijalkowski, K. M.; Gould, C.; Brunner, K.; Molenkamp, L. W. Epitaxy and Structural Properties of (V,Bi,Sb)$_2$Te$_3$ Layers Exhibiting the Quantum Anomalous Hall Effect. *Phys. Rev. Mater.* **2017,** 1, 011201.

36. Chang, C. Z.; Zhao, W. W.; Kim, D. Y.; Wei, P.; Jain, J. K.; Liu, C. X.; Chan, M. H. W.; Moodera, J. S. Zero-Field Dissipationless Chiral Edge Transport and the Nature of Dissipation in the Quantum Anomalous Hall State. *Phys. Rev. Lett.* **2015,** 115, 057206.

37. Chang, C. Z.; Zhao, W. W.; Li, J.; Jain, J. K.; Liu, C. X.; Moodera, J. S.; Chan, M. H. W. Observation of the Quantum Anomalous Hall Insulator to Anderson Insulator Quantum Phase Transition and its Scaling Behavior. *Phys. Rev. Lett.* **2016,** 117, 126802.

38. Bestwick, A. J.; Fox, E. J.; Kou, X. F.; Pan, L.; Wang, K. L.; Goldhaber-Gordon, D. Precise Quantization of the Anomalous Hall Effect near Zero Magnetic Field. *Phys. Rev. Lett.* **2015,** 114, 187201.

39. Grauer, S.; Schreyeck, S.; Winnerlein, M.; Brunner, K.; Gould, C.; Molenkamp, L. W. Coincidence of Superparamagnetism and Perfect Quantization in the Quantum Anomalous Hall State. *Phys. Rev. B* **2015,** 92, 201304.

40. Zhou, L.-J.; Mei, R.; Zhao, Y.-F.; Zhang, R.; Zhuo, D.; Yan, Z.-J.; Yuan, W.; Kayyalha, M.; Chan, M. H. W.; Liu, C.-X.; Chang, C.-Z. Confinement-Induced Chiral Edge Channel Interaction in Quantum Anomalous Hall Insulators. *arXiv:2207.08371* **2022**.

41. Yuan, W.; Zhou, L.-J.; Yang, K.; Zhao, Y.-F.; Zhang, R.; Yan, Z.; Zhuo, D.; Mei, R.; Chan, M. H. W.; Kayyalha, M.; Liu, C.-X.; Chang, C.-Z. Electrical Switching of the Edge Current Chirality in Quantum Anomalous Hall Insulators. *arXiv:2205.01581* **2022**.

42. Yasuda, K.; Mogi, M.; Yoshimi, R.; Tsukazaki, A.; Takahashi, K. S.; Kawasaki, M.; Kagawa, F.; Tokura, Y. Quantized chiral edge conduction on domain walls of a magnetic topological insulator. *Science* **2017,** 358, 1311-1314.

43. Zhang, J. S.; Chang, C. Z.; Zhang, Z. C.; Wen, J.; Feng, X.; Li, K.; Liu, M. H.; He, K.; Wang, L. L.; Chen, X.; Xue, Q. K.; Ma, X. C.; Wang, Y. Y. Band Structure Engineering in $(Bi_{1-x}Sb_x)_2Te_3$ Ternary Topological Insulators. *Nat. Commun.* **2011,** 2, 574.

44. Kong, D. S.; Chen, Y. L.; Cha, J. J.; Zhang, Q. F.; Analytis, J. G.; Lai, K. J.; Liu, Z. K.; Hong, S. S.; Koski, K. J.; Mo, S. K.; Hussain, Z.; Fisher, I. R.; Shen, Z. X.; Cui, Y. Ambipolar field effect in the ternary topological insulator $(Bi_xSb_{1-x})_2Te_3$ by composition tuning. *Nat. Nanotechnol.* **2011,** 6, 705-709.

45. Li, W.; Claassen, M.; Chang, C. Z.; Moritz, B.; Jia, T.; Zhang, C.; Rebec, S.; Lee, J. J.; Hashimoto, M.; Lu, D. H.; Moore, R. G.; Moodera, J. S.; Devereaux, T. P.; Shen, Z. X. Origin of the Low Critical Observing Temperature of the Quantum Anomalous Hall Effect in V-Doped $(Bi, Sb)_2Te_3$ Film. *Sci. Rep.* **2016,** 6, 32732

46. Kou, X. F.; Pan, L.; Wang, J.; Fan, Y. B.; Choi, E. S.; Lee, W. L.; Nie, T. X.; Murata, K.; Shao, Q. M.; Zhang, S. C.; Wang, K. L. Metal-to-Insulator Switching in Quantum Anomalous Hall States. *Nat. Commun.* **2015,** 6, 8474.

47. Wang, W. B.; Ou, Y. B.; Liu, C.; Wang, Y. Y.; He, K.; Xue, Q. K.; Wu, W. D. Direct Evidence of Ferromagnetism in a Quantum Anomalous Hall System. *Nat. Phys.* **2018,** 14, 791-795.

48. Chang, C. Z.; Zhang, Z. C.; Li, K.; Feng, X.; Zhang, J. S.; Guo, M. H.; Feng, Y.; Wang, J.; Wang, L. L.; Ma, X. C.; Chen, X.; Wang, Y. Y.; He, K.; Xue, Q. K. Simultaneous Electrical-Field-Effect Modulation of Both Top and Bottom Dirac Surface States of Epitaxial Thin Films of Three-Dimensional Topological Insulators. *Nano Lett.* **2015,** 15, 1090-1094.